\documentclass[aps,prd,onecolumn,groupedaddress,nofootinbib]{revtex4}
\usepackage{graphicx}

\begin{document}

\title{CP violation for neutral charmed meson decays to CP eigenstates}
\thanks{Supported in part by National Natural Science Foundation of China}
\author{\bf Dong-sheng Du}
\email{duds@mail.ihep.ac.cn}

\affiliation{Institute of High Energy Physics, Chinese Academy of
Sciences, P.O. Box 918 (4), Beijing 100049, China }

\begin{abstract}
CP asymmetries for neutral charmed meson decays to CP eigenstates
are carefully studied. The formulas and numerical results are
presented. The impact on experiments is briefly discussed.
\end{abstract}

\maketitle

\section{Introduction}
Up to now, we still do not have any experimental evidence for CP
violation in the charm sector. Theoretically, the prediction for
charm mixing is very small. This leads to small CP violating
effects in charm decays. However, searching for large mixing and
CP violation in charm decays is still very interesting not only
for testing the standard model but also for finding new physics(
for a recent review, see Ref.\cite{1}). Because CP eigenstates are
very special, if $D^0-\overline{D^0}$ decay into the same  CP
eigenstates, then the CP violating asymmetry could be enhanced by
interference. So we try to investigate this possibility in detail.
Another advantage of CP eigenstates is that the amplitude ratio
$\overline{A}(\overline{D^0} \to f)/{ A}(D^0 \to f)$ can be
estimated without computing the amplitudes directly. This makes
the computation of the CP asymmetries easier. In this paper, we
shall concentrate on the case of CP eigenstates into which charm
decays.

\section{Time-dependent CP asymmetry}
Define
\begin{eqnarray}
&&CP ~ | D^0 \rangle = | \overline{D^0} \rangle \; , \nonumber \\
&&| D^{}_{\rm S} \rangle = p ~ | D^0\rangle + q ~ | \overline{D^0}
\rangle \; , \nonumber \\
&&| D^{}_{\rm L} \rangle = p ~ | D^0\rangle - q ~ | \overline{D^0}
\rangle \; , \nonumber \\
&&| p |^2 + | q |^2 = 1 \; .
\end{eqnarray}
The corresponding eigenvalues of $| D^{}_{\rm S} \rangle, |
D^{}_{\rm L} \rangle$ are
\begin{eqnarray}
\lambda^{}_{\rm S} = m^{}_{\rm S} - i \frac{\gamma^{}_{\rm S}}{2} \;
, ~~~~~~ \lambda^{}_{\rm L} = m^{}_{\rm L} - i \frac{\gamma^{}_{\rm
L}}{2} \; . \nonumber
\end{eqnarray}
Assuming CPT invariance, the time-evolved states are
\begin{eqnarray}
| D^0_{\rm p}(t) \rangle = g^{}_{+}(t) | D^0 \rangle + \frac{q}{p} ~
g^{}_{-}(t) | \overline{D^0} \rangle \; , \nonumber \\
| \overline{D^0_{\rm p}}(t) \rangle = \frac{p}{q} ~ g^{}_{-}(t) |
D^0 \rangle + g^{}_{+}(t) | \overline{D^0} \rangle \; ,
\end{eqnarray}
where
\begin{equation}
\left\{\begin{array}{lll}
  g^{}_{\pm} & = & \displaystyle \frac{1}{2}\left(e^{-i\lambda^{}_{\rm S}t} \pm e^{-i\lambda^{}_{\rm L}t}\right)\\
  ~ & ~ & ~ \\
  ~ & = & \displaystyle \frac{1}{2} ~ e^{-imt-\frac{\gamma}{2} t} \left\{e^{i \frac{\Delta m}{2}t - \frac{\Delta \gamma}{4}t}
  \pm e^{-i \frac{\Delta m}{2} + \frac{\Delta \gamma}{4}t} \right\} \; ,\\
  ~ & ~ & ~ \\
  \Delta m & = & m^{}_{\rm L} - m^{}_{\rm S} \; , ~~ m = (m^{}_{\rm L} + m^{}_{\rm S})/2 \; ,\\
  ~ & ~ & ~ \\
  \Delta \gamma & = & \gamma^{}_{\rm S} ~ - ~ \gamma^{}_{\rm L} \; , ~~
  \gamma ~ = (\gamma^{}_{\rm L} ~ + ~ \gamma^{}_{\rm S})/2 \; .
\end{array}\right.
\end{equation}
Define the mixing parameter
\begin{equation}
x = \frac{\Delta m}{\gamma} \; , ~~~~~~~~~~~ y = \frac{\Delta
\gamma}{2 \gamma} \; ,
\end{equation}
then the decay amplitudes for final state $f$ are
\begin{eqnarray}
{ A}(D^0_{\rm p}(t) \to f) &=& \langle f | { H}^{}_{\rm eff}
| D^0_{\rm p}(t) \rangle \nonumber \\
&=& g^{}_{+}(t) { A}(f) + \frac{q}{p} ~ g^{}_{-}(t)
\overline{ A}(f) \nonumber \\
&=& { A}(f)\left\{g^{}_{+}(t) + \lambda^{}_{f} g^{}_{-}(t)\right\}
\; ,
\end{eqnarray}
where
\begin{eqnarray}
{ A}(f) &=& \langle f | { H}^{}_{\rm eff} | D^0 \rangle \; ,
\nonumber \\
\overline{{ A}}(f) &=& \langle f | { H}^{}_{\rm eff} |
\overline{D^0} \rangle \; , \nonumber \\
\lambda^{}_{f} &=& \frac{q}{p} ~ \frac{\overline{ A}(f)}{{ A}(f)} \;
.
\end{eqnarray}
Similarly, we put
\begin{eqnarray}
{ A}(\overline{f}) &=& \langle \overline{f} | { H}^{}_{\rm
eff} | D^0 \rangle \; , \nonumber \\
\overline{{ A}}(\overline{f}) &=& \langle \overline{f} | {
H}^{}_{\rm eff} | \overline{D^0} \rangle \; , \nonumber \\
\overline{\lambda}^{}_{\overline{f}} &=& \frac{p}{q} ~ \frac{{
A}(\overline{f})}{\overline{ A}(\overline{f})} \; .
\end{eqnarray}
where $\overline{f}$ is the CP conjugate state of the final state
$f$, and
\begin{eqnarray}
| \overline{f} \rangle = CP ~ | f \rangle = \eta_{\rm CP}(f) | f
\rangle \; , \nonumber
\end{eqnarray}
with $\eta_{\rm CP}(f) = \pm 1$ is the CP eigenvalue (or CP parity).
For the amplitude of $\overline{D^0_{\rm p}}(t) \to \overline{f}$,
we have from Eq. (2)
\begin{eqnarray}
{ A}(\overline{D^0_{\rm p}}(t) \to \overline{f}) &=& \langle
\overline{f} | { H}^{}_{\rm eff}
| \overline{D^0_{\rm p}}(t) \rangle \nonumber \\
&=&  \frac{p}{q} ~ g^{}_{-}(t){ A}(\overline{f}) +
g^{}_{+}(t) \overline{ A}(\overline{f}) \nonumber \\
&=& \overline{ A}(\overline{f})\left\{ g^{}_{+}(t) +
\overline{\lambda}^{}_{\overline{f}} g^{}_{-}(t) \right\} \; .
\end{eqnarray}
Now, it is easy to calculate the time-dependent width
$\Gamma(D^0_{\rm p}(t) \to f)$ and $\Gamma(\overline{D^0_{\rm p}}(t)
\to \overline{f})$. Using Eqs. (3), (5) and (8), we have
\begin{eqnarray}
\Gamma(D^0_{\rm p}(t) \to f) &=& \left|A(D^0_{\rm p}(t) \to
f)\right|^2 \nonumber \\
&=& \left|A(f)\right|^2 \left\{|g^{}_{+}(t)|^2 + 2{\rm
Re}\left[\lambda^{}_f g^*_{+}(t) g^{}_{-}(t)\right] +
|\lambda^{}_f|^2 |g^{}_{-}(t)|^2 \right\} \; , \\
\Gamma(\overline{D^0_{\rm p}}(t) \to \overline{f}) &=&
\left|\overline{ A}(\overline{f})\right|^2 \left\{|g^{}_{+}(t)|^2 +
2{\rm Re}\left[\overline{\lambda}^{}_{\overline {f}} g^*_{+}(t)
g^{}_{-}(t)\right] + |\overline{\lambda}^{}_{\overline{f}}|^2
|g^{}_{-}(t)|^2 \right\} \; .
\end{eqnarray}
\begin{figure}[tbp]
\begin{center}
\includegraphics[width=16cm,height=12cm,angle=0]{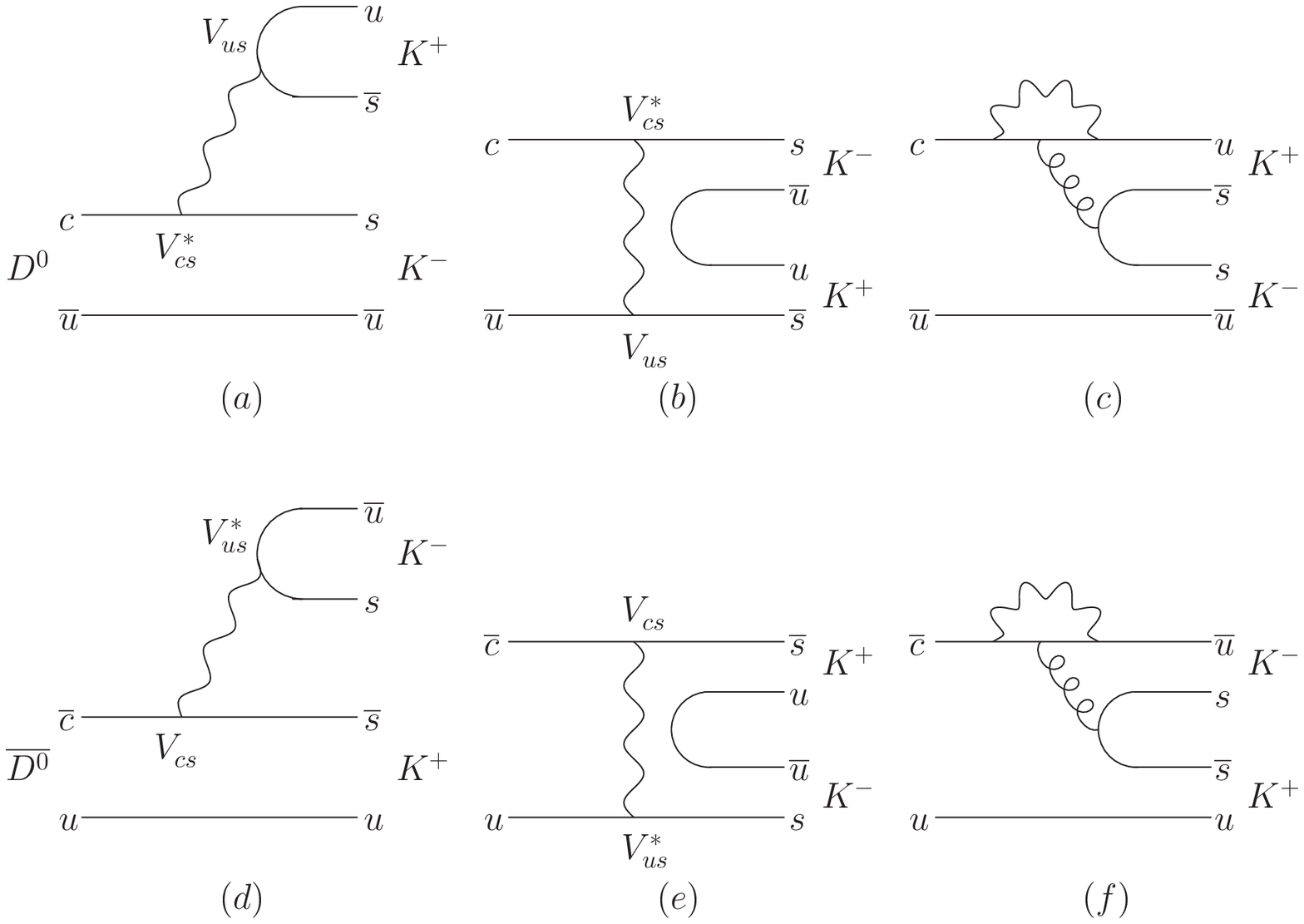}
\end{center}
\caption{ Decay diagrams for $D^0, \overline{D^0} \to K^+ K^-$. }
\end{figure}
In order to compute $\lambda^{}_f$ and
$\overline{\lambda}^{}_{\overline{f}}$, we need first to compute the
amplitude ratios $\overline{A}(f)/A(f)$ and
$A(\overline{f})/\overline{A}(\overline{f})$. As an example, we
consider $D^0, \overline{D^0} \to K^+ K^-$. Draw the decay diagrams
(Fig. 1), we see that if we neglect the penguin diagram
contribution, the $D^0$ and $\overline{D^0}$ decay diagrams involve
only one CKM factor $V^{}_{us} V^*_{cs}$ and $V^*_{us} V^{}_{cs}$,
respectively. The only difference of $D^0$ and $\overline{D^0}$
decay diagrams is that the initial and final particles change into
their CP counterparts. So
\begin{eqnarray}
\frac{\overline{A}(f)}{A(f)} &=& \frac{\overline{A}(\overline{D^0}
\to K^+ K^-)}{A(D^0 \to K^+ K^-)} \nonumber \\
&=& \eta^{}_{\rm CP}(K^+ K^-) \frac{V^*_{us} V^{}_{cs}}{V^{}_{us}
V^*_{cs}} = \eta^{}_{\rm CP}(K^+ K^-) = +1 \; .
\end{eqnarray}
In Eq. (11), $V^{}_{cs}$ and $V^{}_{us}$ are both real in
Wolfenstein parametrization for CKM matrix and $\eta^{}_{\rm CP}(f)$
is the CP parity of the final state $f$. Usually $\eta_{CP}(f) = \pm
1$ for different $f$. Actually, we can prove that (see the appendix
in Ref.\cite{2}) if the decays of $D^0$ and $\overline{D^0}$ only
involve one CKM factor respectively, then the ratio
\begin{equation}
\frac{\overline{A}(f)}{A(f)} = \eta^{}_{\rm CP} (f)\frac{e^{-i
\varphi^{}_{\rm wk}}}{e^{i \varphi^{}_{\rm wk}}} = \eta^{}_{\rm
CP}(f) \; .
\end{equation}
The last equality holds only for charm decay because all the CKM
matrix elements involved are real, if we neglect the penguin
contribution.

Define
\begin{equation}
\rho^{}_f = \frac{\overline{A}(f)}{A(f)} \; , ~~~~~~~
\overline{\rho}^{}_{\overline{f}} =
\frac{A(\overline{f})}{\overline{A}(\overline{f})} \; .
\end{equation}
From Eqs. (6) and (7), we have
\begin{eqnarray}
\lambda^{}_f &=& \frac{q}{p} \rho^{}_f = \eta^{}_{\rm CP}(f)
\left|\frac{q}{p}\right| e^{-i\varphi} \; , \nonumber \\
\overline{\lambda}^{}_{\overline{f}} &=& \frac{p}{q}
\overline{\rho}^{}_{\overline{f}} = \eta^{}_{\rm CP}(f)
\left|\frac{p}{q}\right| e^{i\varphi}\; .
\end{eqnarray}
After a straightforward calculation we arrive at
\begin{eqnarray}
\Gamma(D^0_{\rm p}(t) \to f) &=& \frac{1}{4} e^{-\gamma t} |A(f)|^2
\left\{(1+\left|\frac{q}{p}\right|^2) (e^{-\frac{1}{2}\Delta \gamma
t} + e^{\frac{1}{2} \Delta \gamma t} ) +
2(1-\left|\frac{q}{p}\right|^2) \cos \Delta m t\right. \nonumber
\\
&& \left. + 2\eta^{}_{\rm CP}(f) \left|\frac{q}{p}\right| \left[ (
e^{-\frac{1}{2}\Delta \gamma t} - e^{\frac{1}{2} \Delta \gamma t} )
\cos \varphi + 2 \sin \varphi \sin \Delta m t \right]
\right\} \; , \\
\Gamma(\overline{D^0_{\rm p}}(t) \to f) &=& \frac{1}{4} e^{-\gamma
t} |\overline{A}(\overline{f})|^2 \left\{
(1+\left|\frac{p}{q}\right|^2) (e^{-\frac{1}{2}\Delta \gamma t} +
e^{\frac{1}{2} \Delta \gamma t} ) + 2(1-\left|\frac{p}{q}\right|^2)
\cos \Delta m t\right. \nonumber
\\
&& \left. + 2\eta^{}_{\rm CP}(f) \left|\frac{p}{q}\right| \left[(
e^{-\frac{1}{2}\Delta \gamma t}- e^{\frac{1}{2} \Delta \gamma t})
\cos \varphi - 2 \sin \varphi \sin \Delta m t \right] \large
\right\} \; .
\end{eqnarray}
Assume
\begin{equation}
|A(f)| = |\overline{A}(\overline{f})| \; .
\end{equation}
This is guaranteed by our approximation of neglecting the penguin,
because in that case only one CKM factor appears \cite{3}.

The time-dependent CP asymmetry is
\begin{eqnarray}
C^{}_f(t) &=& \frac{\Gamma(D^0_{\rm p}(t) \to f) -
\Gamma(\overline{D^0_{\rm p}}(t) \to \overline{f})}{\Gamma(D^0_{\rm
p}(t) \to f) + \Gamma(\overline{D^0_{\rm p}}(t) \to \overline{f})}
\equiv
\frac{N^{}_f(t)}{D^{}_f(t)} \; , \\
N^{}_f(t) &=& (\left|\frac{q}{p}\right|^2 -
\left|\frac{p}{q}\right|^2 )(e^{-\frac{1}{2} \Delta \gamma t} +
e^{\frac{1}{2} \Delta \gamma t}) - 2 (\left|\frac{q}{p}\right|^2 -
\left|\frac{p}{q}\right|^2 ) \cos \Delta m t \nonumber \\
&& + 2\eta^{}_{\rm CP}(f) (\left|\frac{q}{p}\right| -
\left|\frac{p}{q}\right|) (e^{-\frac{1}{2} \Delta \gamma t} -
e^{\frac{1}{2} \Delta \gamma t}) \cos \varphi + 4\eta^{}_{\rm CP}(f)
(\left|\frac{q}{p}\right| + \left|\frac{p}{q}\right|) \sin\varphi
\sin \Delta m t \; ,\nonumber \\ \\
D^{}_f(t) &=& (2+\left|\frac{q}{p}\right|^2 +
\left|\frac{p}{q}\right|^2 )(e^{-\frac{1}{2} \Delta \gamma t} +
e^{\frac{1}{2} \Delta \gamma t}) + 2 (2-\left|\frac{q}{p}\right|^2 -
\left|\frac{p}{q}\right|^2 ) \cos \Delta m t \nonumber \\
&& + 2\eta^{}_{\rm CP}(f) (\left|\frac{q}{p}\right| +
\left|\frac{p}{q}\right|) (e^{-\frac{1}{2} \Delta \gamma t} -
e^{\frac{1}{2} \Delta \gamma t}) \cos \varphi + 4\eta^{}_{\rm CP}(f)
(\left|\frac{q}{p}\right| - \left|\frac{p}{q}\right|) \sin\varphi
\sin \Delta m t \; . \nonumber \\
\end{eqnarray}
In Eqs. (19) and (20), there are several parameters we need to know:
the phase $\varphi$, $x = \Delta m/\gamma$, $y = \Delta
\gamma/2\gamma$, $|q|/|p|$, etc. But we do know that $|x| \sim |y|
\lesssim 10^{-2}$(Ref.\cite{1}), and $|q|/|p|$ is very close to
unity. Some people assume \cite{4} that $|q|/|p| - |p|/|q| \lesssim
\pm 1\%$. As for the phase $\varphi$,
\begin{equation}
\frac{q}{p} = \left[\frac{\displaystyle M^*_{12} - \frac{i}{2}
\Gamma^*_{12}}{\displaystyle M^{}_{12} - \frac{i}{2}
\Gamma^{}_{12}}\right]^{1/2} = \left|\frac{q}{p}\right|
e^{-i\varphi} \; .
\end{equation}
In $B^0-\overline{B^0}$ system, the box diagram dominance leads to
\begin{equation}
\left(\frac{q}{p}\right)^{}_{\rm B} \approx
\sqrt{\frac{M^*_{12}}{M^{}_{12}}} \approx e^{-2i\beta} \; , ~~
\varphi_{\rm B} =2\beta \; .
\end{equation}
In charm case, if we can use the same approximation and assume
$b$-quark  plays the dominant role in the corresponding box diagram,
then
\begin{equation}
\left(\frac{q}{p}\right)^{}_{\rm D} \approx e^{-2i\gamma} \; , ~~~
\varphi_{\rm D} = 2\gamma \; .
\end{equation}
But it is not the case for charm. Firstly, the box diagram does not
dominate. Secondly, even in the box diagram because $|V^{}_{ub}|$ is
very small compared with $|V^{}_{ud}|$ and $|V^{}_{us}|$, the
internal $b$-quark contribution may be not important. Furthermore,
$V^{}_{ud}$ and $V^{}_{us}$ do not carry the weak phase,  $\varphi$
can also get the contribution from the $i\Gamma^{}_{12}/2$ term.
Anyway, we do not know the value of $\varphi$, so just keep it as a
free parameter. For $e^{\pm \frac{1}{2}\Delta \gamma t}$, using
$\Delta \gamma /(2\gamma) = y$, we have $e^{\pm \frac{1}{2}\Delta
\gamma t} = e^{\pm y \gamma t} = e^{\pm yt/\tau^{}_{\rm D^0}}$,
because $y \lesssim 10^{-2}$, $e^{\pm y t/\tau^{}_{\rm D^0}}$ is
around unity.
\begin{eqnarray}
N^{}_f(t) &\simeq& -4\eta^{}_{\rm CP}(f) (\left|\frac{q}{p}\right| -
\left|\frac{p}{q}\right|)(y \gamma t) \cos \varphi + 8 \eta^{}_{\rm
CP}(f) \sin \varphi \sin\Delta m t \; , \\
D^{}_f(t) &\simeq& 8 \; , \\
C^{}_f(t) &\approx& \eta^{}_{\rm CP}(f) \left\{\frac{y \gamma t}{2}
(\left|\frac{p}{q}\right| - \left|\frac{q}{p}\right|)\cos \varphi +
\sin\varphi \sin \Delta m t \right\} \nonumber \\
&=& \eta^{}_{\rm CP}(f) \left\{\frac{1}{2} y \gamma t
(\left|\frac{p}{q}\right| - \left|\frac{q}{p}\right|)\cos \varphi +
\sin\varphi \sin (x \gamma t) \right\} \; .
\end{eqnarray}

\section{Time-integrated CP asymmetry}
In order to have more statistics, we integrate the time-dependent
observables with time. We first list some useful quantities:
\begin{eqnarray}
G^{}_+ &=& \int^{\infty}_{0} {\rm d}t ~ |g^{}_+(t)|^2 =
\frac{2+x^2-y^2}{2\gamma (1+x^2)(1-y^2)} \approx \frac{1}{\gamma} \;
, \\
G^{}_- &=& \int^{\infty}_{0} {\rm d}t ~ |g^{}_-(t)|^2 =
\frac{x^2+y^2}{2\gamma (1+x^2)(1-y^2)} \approx
\frac{x^2+y^2}{2\gamma} \; , \\
G^{}_{+-} &=& \int^{\infty}_{0} {\rm d}t ~ g^{*}_+(t) g^{}_-(t)=
\frac{-y(1+x^2)+ix(1-y^2)}{2\gamma (1+x^2)(1-y^2)} \approx
\frac{-y+ix}{2\gamma} \; ,
\end{eqnarray}
for $x^2, y^2 \ll 1$. It is straightforward to get the integrated
decay width. From Eqs. (9) and (10) we have
\begin{eqnarray}
\Gamma(D^0_{\rm p} \to f) = \int^{\infty}_{0} {\rm d}t ~
\Gamma(D^0_{\rm p}(t) \to f) = |A(f)|^2 \left\{G^{}_+ + 2{\rm
Re}(\lambda^{}_f G^{}_{+-}) + |\lambda^{}_f|^2 G^{}_-\right\} \; ,
\\
\Gamma(\overline{D^0_{\rm p}} \to \overline{f}) = \int^{\infty}_{0}
{\rm d}t ~ \Gamma(\overline{D^0_{\rm p}}(t) \to \overline{f}) =
|\overline{A}(\overline{f})|^2 \left\{G^{}_+ + 2{\rm
Re}(\overline{\lambda}^{}_{\overline{f}} G^{}_{+-}) +
|\overline{\lambda}^{}_{\overline{f}}|^2 G^{}_-\right\} \; .
\end{eqnarray}
Again we assume $|A(f)|=|\overline{A}(\overline{f})|$, then the
time-integrated CP asymmetry is
\begin{eqnarray}
C^{}_f&=& \frac{\Gamma(D^0_{\rm p} \to f) -
\Gamma(\overline{D^0_{\rm p}} \to \overline{f})}{\Gamma(D^0_{\rm p}
\to f) + \Gamma(\overline{D^0_{\rm p}} \to \overline{f})} \equiv
\frac{N^{}_f}{D^{}_f} \; , \\
N^{}_f &=& -2\eta^{}_{\rm CP}(f) \left[y(\left|\frac{q}{p}\right| -
\left|\frac{p}{q}\right|) \cos \varphi -
x(\left|\frac{q}{p}\right|+\left|\frac{p}{q}\right|) \sin \varphi
\right] + (x^2+y^2) (\left|\frac{q}{p}\right|^2 -
\left|\frac{p}{q}\right|^2
) \; , \\
D^{}_f &=& 4 - 2\eta^{}_{\rm CP}(f) \left[y(\left|\frac{q}{p}\right|
+ \left|\frac{p}{q}\right|) \cos \varphi +
x(\left|\frac{p}{q}\right|-\left|\frac{q}{p}\right|) \sin
\varphi\right] + (x^2+y^2) (\left|\frac{q}{p}\right|^2 +
\left|\frac{p}{q}\right|^2 ) \nonumber \\
&\approx& 4 \; .
\end{eqnarray}
Finally, neglecting the $(x^2+y^2)$ term in Eq. (33), one can obtain
\begin{equation}
C^{}_f \simeq \eta^{}_{\rm CP}(f) \left\{-\frac{y}{2}
(\left|\frac{q}{p}\right|-\left|\frac{p}{q}\right|) \cos \varphi +
x\sin \varphi \right\}\; . 
\end{equation}
In Ref.\cite{3}, the first term in Eq. (35) is omitted and the CP
paritIy factor $\eta _{\rm CP}(f)$ is missing.

Up to now, we have only discussed incoherent $D^0-\overline{D^0}$
decays. Sometimes $D^0-\overline{D^0}$ pairs are produced
coherently, such as in $e^+ e^-$ colliding machines (BES and
CLEO-c).

The time-evolved coherent state of $D^0-\overline{D^0}$ pair can be
written as \cite{3}
\begin{equation}
|i\rangle = |D^0(k^{}_1, t^{}_1) \overline{D^0}(k^{}_2, t^{}_2)
\rangle + \eta |\overline{D^0}(k^{}_1, t^{}_1) D^0(k^{}_2, t^{}_2)
\rangle \; ,
\end{equation}
where $\eta$ is the charge conjugation parity or orbital angular
momentum parity of the $D^0- \overline{D^0}$ pair.

Because $D^0 \to l^+ X$ and $\overline{D^0} \to l^- X$ only, we can
use the semileptonic decay to tagg one of the two time-evolved
states $D^0_{\rm p}(t)$ and $\overline{D^0_{\rm p}}(t)$. We define
the leptonic-tagging CP asymmetry $C^{}_{fl}$ as
\begin{equation}
C^{}_{fl} = \frac{N(l^-, f) - N(l^+, \overline{f})}{N(l^-, f)
+N(l^+, \overline{f})} \; ,
\end{equation}
where
\begin{equation}
N(l^-, f) = \int^{\infty}_{0}{\rm d}t^{}_1 {\rm d}t^{}_2 ~ |\langle
l^-, f|H^{}_{\rm eff}|i\rangle|^2
\end{equation}
is proportional to the number of events in which $D^0_{\rm p}(k, t)
\to l^- X$ as tagging in one side and the other side is the decay
$\overline{D^0_{\rm p}}(k, t) \to f$ or vice versa. Similarly,
\begin{equation}
N(l^+, \overline{f}) = \int^{\infty}_{0}{\rm d}t^{}_1 {\rm d}t^{}_2
~ |\langle l^+, \overline{f}|H^{}_{\rm eff}|i\rangle|^2 \; .
\end{equation}
Assuming $|A(f)|=|\overline{A}(\overline{f})|$ and
$|A(l^+)|=|\overline{A}(l^-)|$, after a tedious calculation, we have
\begin{eqnarray}
N(l^-, f) &=& |\overline{A}(l^-)A(f)|^2 \left\{G^2_+ + G^2_- +
2|\lambda^{}_f|^2 [G^{}_+ G^{}_- + \eta |G^{}_{+-}|^2]
\right.\nonumber \\
&& \left.+ 2(1+\eta) G^{}_- {\rm Re}(\lambda^{}_f G^{*}_{+-}) +
2(1+\eta) G^{}_+ {\rm Re}(\lambda^{}_f G^{}_{+-}) + 2\eta
{\rm Re}(G^2_{+-}) \right\} \; , \\
N(l^+, \overline{f}) &=& |A(l^+)\overline{A}(\overline{f})|^2
\left\{G^2_+ + G^2_- + 2|\overline{\lambda}^{}_{\overline{f}}|^2
[G^{}_+ G^{}_- + \eta |G^{}_{+-}|^2]
\right.\nonumber \\
&& \left.+ 2(1+\eta) G^{}_- {\rm
Re}(\overline{\lambda}^{}_{\overline{f}} G^{*}_{+-}) + 2(1+\eta)
G^{}_+ {\rm Re}(\overline{\lambda}^{}_{\overline{f}} G^{}_{+-}) +
2\eta {\rm Re}(G^2_{+-}) \right\} \; , \\
C^{}_{fl} &=& \frac{N(l^-, f) - N(l^+, \overline{f})}{N(l^-, f)
+N(l^+, \overline{f})} \equiv \frac{N^{}_{fl}}{D^{}_{fl}}\; , \\
N^{}_{fl} &=& (2+\eta) \frac{x^2+y^2}{2\gamma^2}
(\left|\frac{q}{p}\right|^2-\left|\frac{p}{q}\right|^2) +
(1+\eta)\eta^{}_{\rm CP}(f)\frac{x^2+y^2} {2\gamma^2}
\left[y(\left|\frac{p}{q}\right|-\left|\frac{q}{p}\right|) \cos
\varphi \right.\nonumber \\
&& \left.- x(\left|\frac{q}{p}\right|+\left|\frac{p}{q}\right|) \sin
\varphi \right] + (1+\eta) \eta^{}_{\rm CP}(f) \frac{1}{\gamma^2}
\left[y(\left|\frac{p}{q}\right|-\left|\frac{q}{p}\right|) \cos
\varphi \right.\nonumber \\
&& \left.+ x(\left|\frac{q}{p}\right|+\left|\frac{p}{q}\right|) \sin
\varphi \right] \nonumber \\
&\approx& \frac{2(1+\eta)\eta^{}_{\rm CP}(f)}{\gamma^2}
\left[-\frac{y}{2}(\left|\frac{q}{p}\right|-\left|\frac{p}{q}\right|)
\cos \varphi + x \sin \varphi \right] \; , \\
D^{}_{fl} &=& \frac{2}{\gamma^2} + \frac{(x^2+y^2)^2}{2\gamma^2} +
\frac{1}{2\gamma^2}(\left|\frac{p}{q}\right|+\left|\frac{q}{p}\right|)(2+\eta)
(x^2+y^2) \nonumber \\
&& + (1+\eta)\eta^{}_{\rm CP}(f) \frac{x^2+y^2}{2\gamma^2}
\left[-y(\left|\frac{p}{q}\right|+\left|\frac{q}{p}\right|) \cos
\varphi  - x(\left|\frac{q}{p}\right|-\left|\frac{p}{q}\right|) \sin
\varphi \right]\nonumber \\
&& + (1+\eta)\eta^{}_{\rm CP}(f) \frac{1}{\gamma^2}
\left[-y(\left|\frac{q}{p}\right|+\left|\frac{p}{q}\right|) \cos
\varphi  - x(\left|\frac{q}{p}\right|-\left|\frac{p}{q}\right|) \sin
\varphi \right] + 4\eta \frac{y^2-x^2}{4\gamma^2} \nonumber \\
&\approx& \frac{2}{\gamma^2} \; .
\end{eqnarray}
Finally
\begin{equation}
C^{}_{fl} = \frac{N^{}_{fl}}{N^{}_{fl}} = (1+\eta) \eta^{}_{\rm
CP}(f)
\left\{-\frac{y}{2}(\left|\frac{q}{p}\right|-\left|\frac{p}{q}\right|)
\cos \varphi + x \sin \varphi \right\} \; .
\end{equation}
Comparing Eq. (45) with Eq. (35), we find that $C^{}_{fl}$ is just
twice as large as $C^{}_f$ when the charge conjugation parity or
the orbital angular momentum $l$ is even. This is surprising. From
Eq. (35), the order of magnitude of $C^{}_f$ is $\lesssim
10^{-3}$, because $x\sim y \lesssim 10^{-2}$. Now we present
$C^{}_f$ (theory), $C^{}_f$ (exp.), branching fractions for $D^0,
\overline{D^0}$ decay into CP eigenstates and the number of
$D-\overline{D}$ pairs needed for testing CP asymmetry for
$1\sigma$ signal lower bound in TABLE I, where for the branching
ratios we take most of them from the 2006 particle data booklet
\cite{5}. For ${K^*}^+ {K^*}^-$, $\rho^+ \rho^-$ and $\rho^0
\rho^0$ final states, we use the BSW theoretical estimation
\cite{6}. For the measured CP asymmetries listed in Table I  are
also taken from the 2006 particle data booklet\cite{5}. We use the
formula for $N^{}_{\rm D\overline{D}}$
\begin{eqnarray}
N^{}_{\rm D\overline{D}} &=& \frac{1}{BC^2_{f}} ~~~~~~~~{\rm for }~
1\sigma ~ {\rm signature} \; ; \nonumber \\
N^{}_{\rm D\overline{D}} &=& \frac{9}{BC^2_{f}} ~~~~~~~~{\rm for }~
3\sigma ~ {\rm signature} \; . \nonumber
\end{eqnarray}
\begin{table}
\caption{The number of $D \overline{D}$ pairs needed for testing CP
asymmetry} \vspace{0.3cm}
\begin{center}
\begin{tabular}{ccccc}
\hline \hline
$D^0 \to f$ & $C^{}_f$ (theory) &  $C^{}_f$ (exp.) & BR & $N^{}_{\rm D\overline{D}}$ ($1\sigma$ lower bound)\\
~ & ~ & ~ & ~ & ~ \\
$K^+ K^-$ & ~ & $0.014 \pm 0.010$ & $(3.84\pm0.10)\times 10^{-3}$ & $2.60\times 10^7$\\
$K^{}_s K^{}_s$ & ~ & $-0.23 \pm 0.19$ & $(3.7\pm0.7)\times 10^{-4}$ & $2.70\times 10^8$\\
${K^*}^+ {K^*}^-$ & ~ & ~ & $1.0\times 10^{-2}$(BSW) & $1\times 10^7$\\
$\pi^+ \pi^-$ & $\lesssim 10^{-3}$ & $0.013\pm0.012$~ & $(1.364\pm0.032)\times 10^{-3}$ & $7.4\times 10^7$\\
$\pi^0 \pi^0$ & ~ & $0.00\pm0.05$ & $(7.9\pm0.8)\times 10^{-4}$ & $1.26\times 10^8$\\
$\rho^0 \pi^0$ & ~ & ~ & $(3.2\pm0.4)\times 10^{-3}$ & $3.13\times 10^7$\\
$\rho^+ \rho^-$ & ~ & ~ & $1.3\times 10^{-2}$(BSW) & $7.69\times 10^6$\\
$\rho^0 \rho^0$ & ~ & ~ & $1.2\times 10^{-3}$ (BSW)& $8.33\times 10^7$\\
$\phi \pi^0$ & ~ & ~ & $(7.4\pm0.5)\times 10^{-4}$ & $1.35\times 10^8$\\
$\phi \eta$ & ~ & ~ & $(1.4\pm0.4)\times 10^{-4}$ & $7.14\times 10^8$\\
${K^*}^0 {K^*}^0$ & ~ & ~ & $(7\pm5)\times 10^{-5}$ & $1.43\times 10^9$\\
\hline \hline
\end{tabular}
\end{center}
\end{table}

\section{Summary and Conclusion}
We have computed the time-dependent and time-integrated CP asymmetry
for neutral charmed meson decays into CP eigenstates. We present CP
asymmetry not only for incoherent $D^0-\overline{D^0}$, but also
coherent $D^0 \overline{D^0}$ pairs. We find that the
time-integrated CP asymmetries are very small (order of $\lesssim
10^{-3}$). We also give the lower bound for the number of
$D\overline{D}$ pairs needed for testing the CP asymmetries. At
present, the integrated luminosities for $e^+ e^-$ colliders are :
\begin{eqnarray}
~~~~~~~~~~~~~~~~~~~~~~~~~~~~~~~~~~~~~~~\begin{array}{ll}
  {\rm BES ~ II:} & 27 ~ {\rm pb}^{-1} \\
  {\rm BES ~ III:} & 20 ~ {\rm fb}^{-1}\\
  {\rm CLEO-c:} & 281 ~ {\rm pb}^{-1}
\end{array} ~~ {\rm for} ~4~ {\rm years ~ data ~ taking } \; .\nonumber
\end{eqnarray}
The corresponding $D\overline{D}$ pairs are
\begin{eqnarray}
\begin{array}{ll}
  {\rm BES ~ II:} & 10^5 \\
  {\rm BES ~ III:} & 10^7\\
  {\rm CLEO-c:} & 10^6
\end{array} \; . \nonumber
\end{eqnarray}
In Table I, for $D \to VV$ decays, only when both vector mesons
are longitudinally polarized the $VV$ final states are CP eigen
states. For the corresponding branching ratios in Table I,we
assume that the $ VV$ final states for which both $V$ are
longitudinally polarized  doninate. From Table I we see that the
only hope is relying on BES III and $B$ factories. At
$B$-factories, because the large data sample of charmed meson,
both time-dependent asymmetry and time-integrated asymmetry can be
measured. While for BES III, only time-integrated CP asymmetry can
be tested. Of course, if there is new physics, some surprise may
happen. We can also see from Table I that all the measured CP
asymmetries are consistent with zero.

\acknowledgments{I thank Hai-bo Li, Cai-dian L\"{u}, Mao-zhi Yang
and Zhi-zhong Xing for discussions. This work is supported in part
by the National Natural Science Foundation of China under grants
NSFC 90103011, 10375073 and 90403024.}

\end{document}